\documentclass{IEEEtran}

\usepackage{amsmath,amssymb,amsfonts}
\usepackage{physics}
\usepackage{algorithmic}
\usepackage{graphicx}
\usepackage{eso-pic}
\usepackage{textcomp}
\usepackage{hyperref}
\usepackage[capitalise]{cleveref}
\usepackage{siunitx}
\usepackage[numbers]{natbib}

\usepackage{tikz}
\usepackage{pgfplots}
\pgfplotsset{compat=newest}
\usetikzlibrary{patterns, arrows.meta, spy, shapes.geometric, calc, shapes, positioning, backgrounds}

\usepgfplotslibrary{units}
\pgfplotsset{
    unit code/.code 2 args={\si{#1#2}},
    unit marking pre={(},
    unit marking post={)},
}

\usepackage{todonotes}
\setuptodonotes{inline}

\usepackage{preamble/tikzstyles}
\usepackage{preamble/math}

\def\BibTeX{{\rm B\kern-.05em{\sc i\kern-.025em b}\kern-.08em
    T\kern-.1667em\lower.7ex\hbox{E}\kern-.125emX}}

\begin{document}

\title{Subdivision-based isogeometric analysis for axisymmetric electromagnetic problems}
\author{Devin Balian, Sebastian Schöps and Melina Merkel
    \thanks{
        This work is supported by the joint DFG/FWF Collaborative Research Centre CREATOR
        (DFG: Project-ID 492661287/TRR 361; FWF: 10.55776/F90) at TU Darmstadt, TU Graz and JKU Linz
        and the Graduate School of Computational Engineering at TU Darmstadt.
    }
    \thanks{
        Devin Balian, Sebastian Schöps and Melina Merkel
        are with the Institute for Accelerator Science and Electromagnetic Fields (TEMF) and the
        Graduate School of CE, TU Darmstadt, 64289 Darmstadt, Germany.
        (e-mail: \href{mailto:devin.balian@tu-darmstadt.de}{devin.balian@tu-darmstadt.de},
        \href{mailto:sebastian.schoeps@tu-darmstadt.de}{sebastian.schoeps@tu-darmstadt.de},
        \href{mailto:melina.merkel@tu-darmstadt.de}{melina.merkel@tu-darmstadt.de}).
    }
}

\AddToShipoutPicture*{
\footnotesize\sffamily\raisebox{0.8cm}{
\hspace{1.4cm}\fbox{
\parbox{\textwidth}{
    This work has been submitted to the IEEE for possible publication.
    Copyright may be transferred without notice, after which this version may no longer be accessible.
}}}}

\maketitle

\begin{abstract}
This paper applies a subdivision-based isogeometric method to solve the axisymmetric Maxwell eigenvalue problem.
The reduction to an $H^1$-formulation allows to use a Catmull-Clark construction for both geometry and field discretization.
The approach yields a numerical solution for the electric field, which is $C^1$-continuous everywhere except at extraordinary vertices.
This is demonstrated by computing the eigenmodes of a TESLA 9-cell cavity,
showing smoother fields with less numerical noise than conventional methods.
The convergence rate of the method is numerically analyzed and is in agreement with rates observed in the literature.
\end{abstract}


\section{Introduction}
\label{sec:introduction}
\IEEEPARstart{S}{ubdivision} surfaces have emerged as a promising alternative to standard CAD technologies in recent years.
They overcome the topological limitations of multi-patch NURBS surfaces, while still providing a smooth surface representation.
Besides being a powerful tool for geometric modeling, subdivision methods naturally extend to application in isogeometric analysis (IGA).
First showcased in \cite{Cirak_2000aa}, the Loop subdivision scheme was utilized
for discretization of the Kirchhoff-Love thin-shell formulation,
satisfying the regularity requirements.
While electromagnetic problems typically only require continuity,
the integration of IGA with subdivision surfaces may still be advantageous for the simulation of various problems,
in particular where smoothness of the numerical solution for the electromagnetic fields is a desirable property,
e.g. particle tracking.

Most relevant electromagnetic formulations require not just the scalar space $\Hgrad$,
but also the vectorial spaces $\Hcurl$ and $\Hdiv$ (or variants of them) of a de Rham sequence.
These spaces need special treatment when it comes to discretization, requiring vectorial edge and face basis functions, respectively \cite{Monk_2003aa}.
The classical subdivision schemes (e.g. Catmull-Clark \cite{Catmull_1978aa}, Loop \cite{Loop_1987aa}) however generate scalar $\Hgrad$ limit functions,
which are not applicable for a consistent $\Hcurl$ discretization.

There are specialized edge subdivision schemes in the literature, first introduced by \citeauthor{Wang_2006aa}~\cite{Wang_2006aa},
trying to generalize subdivision to the discrete exterior calculus setting, by enforcing commutativity between subdivision and the exterior derivatives.
The proposed edge subdivision schemes, based on the Loop and Catmull-Clark methods, are further investigated in \cite{De-Goes_2016aa},
where application in the context of numerical analysis, in particular for IGA, is investigated.
Recently, \citeauthor{Piel_2026aa} \cite{Piel_2026aa} provide a rigorous mathematical framework for the generalization to $k$-form subdivision schemes,
and even apply the proposed methods to solve a two-dimensional Maxwell eigenvalue problem.
However, the works above do not use the subdivision limit functions for the IGA discretization
and hence do not obtain increased inter-element regularity for the numerical solution.
\footnote{\citeauthor{Piel_2026aa} call the increased smoothness of their subdivision $k$-forms \emph{subdivision-induced regularity},
which they distinguish from Sobolev regularity.}
To the authors' best knowledge it is still unknown, how the limit functions from these subdivision schemes perform and how to evaluate them exactly.

A second challenge is the extension to three-dimensional problems.
An isogeometric approach using the boundary element method (BEM)
is generally well suited for the application of the subdivision limit functions,
since discretization of the boundary data requires only bivariate basis functions, which subdivision surfaces naturally provide.
This approach is for example investigated in \cite{Li_2016aa, Liu_2018ad}.
A volumetric discretization by the finite element method (FEM) however,
requires trivariate basis functions, making subdivision surfaces inapplicable.
There do exist extensions of well known subdivision algorithms to \emph{subdivision solids},
and they have been applied in the context of IGA \cite{Burkhart_2010aa},
but their analytical properties are even less understood than their two-dimensional counterparts \cite{Dietz_2022aa}.
Also, to our knowledge, there exist to this date no generalization of edge subdivision schemes,
required for an $\Hcurl$ discretization, in the trivariate case.

Considering the above mentioned challenges, this paper investigates the following approach:
For the reduction from 3D to 2D, we exploit the symmetry of an axisymmetric problem setup.
This is a reasonable assumption for various problems of practical relevance, such as simulation of accelerator components.
For the Maxwell eigenvalue problem, the approach results in a problem formulation in $\HgradAxi$,
which can be discretized with standard subdivision limit functions, without the need for a specialized edge subdivision scheme.

The paper is structured as follows:
First, in \cref{sec:maxwell} the Maxwell eigenvalue problem is formulated
and the simplification under axisymmetry is performed.
In \cref{sec:catmark_subd} the basics of Catmull-Clark subdivision,
and especially the construction of the basis functions, are introduced.
Discretization of the problem formulation using subdivision IGA is then performed in \cref{sec:iga}.
All numerical results are presented in \cref{sec:results}.

\section{Axisymmetric Maxwell eigenvalue problem}
\label{sec:maxwell}

Macroscopic electromagnetic phenomena are governed by Maxwell's equations.
Under certain physical assumptions, the equations can be simplified to boundary value problems solvable using numerical methods.
The Maxwell eigenvalue problem we consider in this section,
is an important formulation describing electromagnetic wave propagation
in for example cavities and waveguides \cite{Jackson_1998aa}.

Consider a sufficiently smooth domain $\Omega\subset\real^3$,
constant materials $\varepsilon, \mu > 0$ and zero sources.
The three-dimensional time-harmonic Faraday's and Ampère's laws are then given by the two equations
\begin{subequations}
\begin{align}
    \curl\vb{E} &= -i \omega\mu\vb{H} \label{eq:faraday} \\
    \curl\vb{H} &= i \omega\varepsilon\vb{E} \label{eq:ampere},
\end{align}
\end{subequations}
with the eigenfrequency $\omega$ and imaginary unit $i$ \cite[Chapter 7]{Jackson_1998aa}.
Inserting \eqref{eq:ampere} into \eqref{eq:faraday} yields the $\vb{H}$-based Helmholtz wave equation
\begin{equation}
    \curl\curl\vb{H} = \omega^2\mu\varepsilon \vb{H} .
    \label{eq:helmholtz}
\end{equation}
To numerically compute solutions of \eqref{eq:helmholtz}, a variational formulation is typically used,
and for well-posedness of the problem, appropriate boundary conditions must be imposed.
We consider homogeneous Dirichlet and homogeneous Neumann boundary conditions here.
For notational simplicity, the boundary conditions are not explicitly expressed in the function space notation.
The weak $\vb{H}$-formulation of the Maxwell eigenvalue problem for $(\lambda, \vb{H}) \in \real_{>0}\times\Hcurl$ is then given by
\begin{equation}
    \forall \vb{H}'\in\Hcurl:
    (\curl \vb{H}, \curl \vb{H}')
    = \lambda (\vb{H}, \vb{H}'),
    \label{eq:maxwell_ev_weak}
\end{equation}
where $\vb{H}'$ is a \emph{test function} \cite[Chapter 4]{Monk_2003aa}.
We denote by $\lambda = \omega^2 \varepsilon\mu$ the eigenvalue
and by $(\cdot,\cdot)$ the inner product on $L^2(\Omega)$ and $(L^2(\Omega))^3$.

The following analysis is restricted to the axisymmetric eigenvalue problem.
In particular we assume that the domain $\Omega$ is a solid of revolution of an axisymmetric slice
\begin{equation}
    S := \qty{ (\rho,z) \in\real_{\geq 0}\times\real : (\rho,0,z)\in\Omega } .
\end{equation}
Under this assumption the derivatives $\partial_\varphi$ vanish and formulation \eqref{eq:maxwell_ev_weak} simplifies significantly.
The above function spaces are then to be replaced by two-dimensional suitably weighted versions of them (indicated by a $\rho$-subscript).
Considering magnetic fields with only a $\varphi$-component,
i.e. $\vb H = u \vb{e}_\varphi$, the formulation can be reduced to
\begin{equation}
    \forall v \in\HgradAxi[S]:
    (\rot u, \rot v)_\rho = \lambda (u, v)_\rho,
    \label{eq:maxwell_ev_weak_axi}
\end{equation}
where $\rot: \HgradAxi[S]\to\HdivAxi[S]$ is the \emph{rotated gradient} and
\begin{equation}
    (u,v)_\rho = \iint_S u v \,\rho\dd{\rho}\dd{z}
\end{equation}
is the two-dimensional $\rho$-weighted inner product on $\LtwoAxi[S]$ and $\qty(\LtwoAxi[S])^2$.

For any given eigenpair $(\lambda,u)$, the electric field can be computed by Ampère's law as
\begin{equation}
    \HrotAxi[S]\ni\vb{E} = \frac{-i}{\omega\varepsilon} \rot u,
\end{equation}
with $\rotScal: \HrotAxi[S]\to\LtwoAxi[S]$ being the \emph{scalar curl}.

Note that this ansatz can be generalized by expanding the solution in terms of Fourier modes, see e.g. \cite{Simona_2020aa}.
For simplicity, this is not investigated here.

\section{Catmull-Clark subdivision}
\label{sec:catmark_subd}
There exists a variety of different subdivision algorithms, all with different geometric and numerical properties.
For a general introduction to subdivision, we refer to \cite{Peters_2008aa}.
One of the most well known and conceptionally easy to understand subdivision schemes is \emph{Catmull-Clark} subdivision.
A Catmull-Clark subdivision surface is an almost-everywhere $C^2$ surface, defined on a possibly unstructured quadrilateral mesh.
Restricted to tensor-product patches, the surface coincides with classic cubic B-spline surfaces \cite[Chapter 6.1]{Peters_2008aa}.

The Catmull-Clark subdivision algorithm is originally formulated in \cite{Catmull_1978aa}
and is understood as a refinement algorithm of a coarse control mesh.
The iterative application of the subdivision stencil to the initial mesh will then, in the limit, converge to a smooth surface.
Similar to the well-known B-spline surfaces, this limit surface can also be represented by a linear combination of basis functions
with the initial control points. This point of view is crucial for the application of subdivision to isogeometric analysis.

To define the Catmull-Clark basis functions, first consider the parametric domain
\begin{equation}
    \widehat{S} := [0,1]^2 \times \qty{1,\dots,N}
\end{equation}
of $N$ cells.
All cells are equipped with adjacency information, by topologically identifying the corners and edges of neighboring cells.
The mesh defined by these neighborhood relations is allowed to be unstructured.
In that case, the valence (number of outgoing edges) of a mesh vertex is in the following denoted by $n$
and equals $4$ for regular interior vertices. Vertices of any other valence are called \emph{extraordinary}.

Similar to classical B-spline theory, the subdivision basis functions on each cell $\Sigma\in\widehat{S}$
have support extending onto the neighboring cells.
While these basis functions are defined by the limit of the subdivision process,
they can also be evaluated parametrically, as shown in \cite{Stam_1998aa}.
In the following, for the construction of a parametric representation of the basis functions,
we need to differentiate between \emph{regular} and \emph{irregular} cells,
i.e. cells without or with an extraordinary vertex, respectively.
Cells at the boundary of the mesh also need special treatment
and are usually handled by extrapolation of control points.
We do not discuss this case here; details can be found in \cite{Lacewell_2007aa}.

\subsection{Local regular basis functions}
\begin{figure}[ht]
    \centering
    \begin{tikzpicture}[scale=1.3]

\draw[pattern=north east lines, pattern color=TUDa-3c, opacity=0.5] (1,1) rectangle (2,2);
\draw[-{Latex[length=3pt]}] (1.1,1.1) -- (1.6,1.1) node[right,yshift=0.05cm] {\small $\xi$};
\draw[-{Latex[length=3pt]}] (1.1,1.1) -- (1.1,1.6) node[above] {\small $\eta$};

\draw (0,0) grid (3,3);

\foreach \x in {0,...,3} {
    \foreach \y in {0,...,3} {
        \filldraw (\x,\y) circle (1pt);
    }
}

\pgfplotsset{
    every axis post/.style={
        x=1cm,
        y=0.2cm,
        axis lines=none,
        no marks,
        xmin=0,
        xmax=1,
        domain=0:1
    }
}
\pgfplotscreateplotcyclelist{basis plot}{
    [indices of colormap={0,1,2,3} of TUDa-b]
}

\begin{axis}[at={(0cm,3cm)}, cycle list name=basis plot, cycle list shift=3]
	\addplot+[dashed, opacity=0.5] {1 - 3*x + 3*x^2 - x^3};
	\addplot+[dashed, opacity=0.5] {4 - 6*x^2 + 3*x^3};
	\addplot+[dashed, opacity=0.5] {1 + 3*x + 3*x^2 - 3*x^3};
	\addplot+[dashed, opacity=0.5] {x^3};
\end{axis}
\begin{axis}[at={(1cm,3cm)}, cycle list name=basis plot]
	\addplot {1 - 3*x + 3*x^2 - x^3};
	\addplot {4 - 6*x^2 + 3*x^3};
	\addplot {1 + 3*x + 3*x^2 - 3*x^3};
	\addplot {x^3};
\end{axis}
\begin{axis}[at={(2cm,3cm)}, cycle list name=basis plot, cycle list shift=1]
    \addplot+[dashed, opacity=0.5] {1 - 3*x + 3*x^2 - x^3};
	\addplot+[dashed, opacity=0.5] {4 - 6*x^2 + 3*x^3};
	\addplot+[dashed, opacity=0.5] {1 + 3*x + 3*x^2 - 3*x^3};
	\addplot+[dashed, opacity=0.5] {x^3};
\end{axis}

\begin{axis}[at={(-1cm,0cm)}, rotate=90, cycle list name=basis plot, cycle list shift=3]
	\addplot+[dashed, opacity=0.5] {1 - 3*x + 3*x^2 - x^3};
	\addplot+[dashed, opacity=0.5] {4 - 6*x^2 + 3*x^3};
	\addplot+[dashed, opacity=0.5] {1 + 3*x + 3*x^2 - 3*x^3};
	\addplot+[dashed, opacity=0.5] {x^3};
\end{axis}
\begin{axis}[at={(-1cm,1cm)}, rotate=90, cycle list name=basis plot]
	\addplot {1 - 3*x + 3*x^2 - x^3};
	\addplot {4 - 6*x^2 + 3*x^3};
	\addplot {1 + 3*x + 3*x^2 - 3*x^3};
	\addplot {x^3};
\end{axis}
\begin{axis}[at={(-1cm,2cm)}, rotate=90, cycle list name=basis plot, cycle list shift=1]
	\addplot+[dashed, opacity=0.5] {1 - 3*x + 3*x^2 - x^3};
	\addplot+[dashed, opacity=0.5] {4 - 6*x^2 + 3*x^3};
	\addplot+[dashed, opacity=0.5] {1 + 3*x + 3*x^2 - 3*x^3};
	\addplot+[dashed, opacity=0.5] {x^3};
\end{axis}

\end{tikzpicture}
    \caption{
        Schematic of a regular cell and its structured neighborhood.
        The four univariate cubic B-splines are depicted for both parametric directions.
        Based on \cite{Liu_2020ab}.
    }
    \label{fig:cell_regular}
\end{figure}
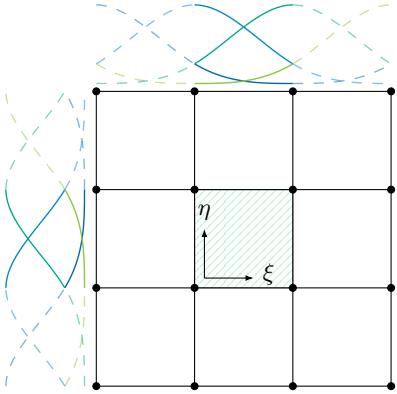

For a cell without an extraordinary vertex, consider the four uniform cubic B-splines $N_i: [0,1]\to\real$, defined by
\begin{align*}
    N_1(\xi) &= \frac{1}{6} \left(1 - 3\xi + 3\xi^2 - \xi^3 \right) \\
    N_2(\xi) &= \frac{1}{6} \left(4 - 6\xi^2 + 3\xi^3 \right) \\
    N_3(\xi) &= \frac{1}{6} \left(1 + 3\xi + 3\xi^2 - 3\xi^3 \right) \\
    N_4(\xi) &= \frac{1}{6} \xi^3.
\end{align*}
The Catmull-Clark basis functions on this regular cell then coincide with the $4\times4$ tensor-product B-splines
\begin{equation}
    \widehat{b}_{ij}: [0,1]^2 \to\real
    \quad
    (\xi, \eta) \mapsto N_i(\xi) N_j(\eta) .
\end{equation}
On each regular cell $\Sigma$, these local basis functions are typically sorted lexicographically
and denoted with a single linear index as $\widehat{b}_{\Sigma,i}$.
Each of these basis functions is associated with a vertex in the $3\times3$ grid of neighboring cells.
The \num{16} vertices of this grid and the basis functions along both parametric directions
are schematically depicted in \cref{fig:cell_regular}.

\subsection{Local irregular basis functions}
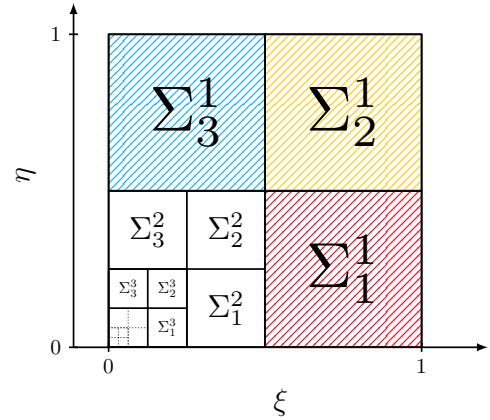
\begin{figure}[ht]
    \centering
    \begin{tikzpicture}[scale=0.8]
\begin{axis}[
    xmin=0, xmax=1.1,
    ymin=0, ymax=1.1,
    xtick={0,1},
    ytick={0,1},
    xlabel={$\xi$},
    ylabel={$\eta$},
    label style={font=\Large},
    axis lines=left,
    axis equal,
    axis line style={-latex},
    thick,
    every tick/.append style={thick, black}
]

\draw[pattern=north east lines, pattern color=TUDa-2b] (0,0.5) rectangle (0.5,1);
\draw[pattern=north east lines, pattern color=TUDa-6b] (0.5,0.5) rectangle (1,1);
\draw[pattern=north east lines, pattern color=TUDa-9b] (0.5,0) rectangle (1,0.5);

\draw[step=0.5] (0,0) grid (1,1);
\draw[step=0.25] (0,0) grid (0.5,0.5);
\draw[step=0.125] (0,0) grid (0.25,0.25);
\draw[step=0.0625, densely dotted, thin] (0,0) grid (0.125,0.125);
\draw[step=0.03125, densely dotted, thin] (0,0) grid (0.0625,0.0625);

\node[scale=3] at (0.75,0.25) {$\Sigma^1_1$};
\node[scale=3] at (0.75,0.75) {$\Sigma^1_2$};
\node[scale=3] at (0.25,0.75) {$\Sigma^1_3$};

\node[scale=1.5] at (0.75/2,0.25/2) {$\Sigma^2_1$};
\node[scale=1.5] at (0.75/2,0.75/2) {$\Sigma^2_2$};
\node[scale=1.5] at (0.25/2,0.75/2) {$\Sigma^2_3$};

\node[scale=0.75] at (0.75/4,0.25/4) {$\Sigma^3_1$};
\node[scale=0.75] at (0.75/4,0.75/4) {$\Sigma^3_2$};
\node[scale=0.75] at (0.25/4,0.75/4) {$\Sigma^3_3$};

\end{axis}
\end{tikzpicture}
    \caption{
        Partition of the unit square with infinitely many tiles.
        Based on \cite{Stam_1998aa}.
    }
    \label{fig:subd_unit_square}
\end{figure}

Handling irregular cells is more involved than their regular counterpart,
and lies at the heart of each subdivision scheme.
Because for such an irregular region, no structured $3\times3$ grid of cells can be found,
the tensor-product B-splines cannot meaningfully be defined.
Instead, the idea is to subdivide repeatedly, until the parameters $(\xi, \eta)$ lie inside a regular cell
and then evaluate the regular bi-cubic B-splines.
Mathematically, this is achieved as follows:
Consider an infinite family of tiles $\Sigma_k^\ell \subset [0,1]^2$ partitioning the unit square, as depicted in \cref{fig:subd_unit_square}.
The upper index $\ell > 0$ denotes the level of subdivision
and the lower index $k \in\qty{1,2,3}$ denotes the position inside the refined square (bottom right, top right, top left).
For $(\xi, \eta) \in\Sigma_k^\ell$ consider the transformations $t_k^\ell: \Sigma_k^\ell \to [0,1]^2$, defined by
\begin{align*}
    t_1^\ell(\xi, \eta) &= (2^\ell \xi - 1, 2^\ell \eta) \\
    t_2^\ell(\xi, \eta) &= (2^\ell \xi - 1, 2^\ell \eta - 1) \\
    t_3^\ell(\xi, \eta) &= (2^\ell \xi, 2^\ell \eta - 1).
\end{align*}
These transformations simply map the parametric coordinates $(\xi, \eta)$ from a refined tile to the standard unit square
in an affine way.
This mapping is required to evaluate the B-splines on each tile.

\begin{figure}[ht]
    \centering
    \begin{tikzpicture}[scale=1.3]

\draw[pattern=north east lines, pattern color=TUDa-2b, opacity=0.5] (0,0.5) rectangle (0.5,1);
\draw[pattern=north east lines, pattern color=TUDa-6b, opacity=0.5] (0.5,0.5) rectangle (1,1);
\draw[pattern=north east lines, pattern color=TUDa-9b, opacity=0.5] (0.5,0) rectangle (1,0.5);

\draw (0,0) grid (2,2);
\draw (0,-1) grid (2,0);
\draw (-1,0) grid (0,2);
\draw (0,0) -- (-0.707,-0.707) -- (-0.707,-1.707) -- (0,-1);
\draw (0,0) -- (-0.707,-0.707) -- (-1.707,-0.707) -- (-1,0);

\foreach \x in {-1,...,2} {
    \foreach \y in {-1,...,2} {
        \ifnum \x > -1
            \filldraw (\x,\y) circle (1pt);
        \fi
        \ifnum \y > -1
            \filldraw (\x,\y) circle (1pt);
        \fi
    }
}
\filldraw (-0.707,-0.707) circle (1pt);
\filldraw (-0.707,-1.707) circle (1pt);
\filldraw (-1.707,-0.707) circle (1pt);

\draw[densely dashed, step=0.5, lightgray] (0,0) grid (1.5,1.5);
\draw[densely dashed, step=0.5, lightgray] (0,-0.5) grid (1.5,0);
\draw[densely dashed, step=0.5, lightgray] (-0.5,0) grid (0,1.5);
\draw[densely dashed, lightgray] (0,0) -- (-0.354,-0.354) -- (-0.354,-0.854) -- (0,-0.5);
\draw[densely dashed, lightgray] (0,0) -- (-0.354,-0.354) -- (-0.854,-0.354) -- (-0.5,0);

\foreach \x in {-1,...,3} {
    \foreach \y in {-1,...,3} {
        \ifnum \x > -1
            \filldraw[lightgray] (\x/2,\y/2) circle (1pt);
        \fi
        \ifnum \y > -1
            \filldraw[lightgray] (\x/2,\y/2) circle (1pt);
        \fi
    }
}

\draw[densely dashed, step=0.5, darkgray!80] (0,0) grid (1,1);
\draw[densely dashed, step=0.5, darkgray!80] (0,-0.5) grid (1,0);
\draw[densely dashed, step=0.5, darkgray!80] (-0.5,0) grid (0,1);
\draw[densely dashed, darkgray!80] (0,0) -- (-0.354,-0.354) -- (-0.354,-0.854) -- (0,-0.5);
\draw[densely dashed, darkgray!80] (0,0) -- (-0.354,-0.354) -- (-0.854,-0.354) -- (-0.5,0);

\draw[thick] (0,0) rectangle (1,1);
\draw[-{Latex[length=3pt]}] (0.1,0.1) -- (0.6,0.1) node[right,yshift=0.05cm] {\small $\xi$};
\draw[-{Latex[length=3pt]}] (0.1,0.1) -- (0.1,0.6) node[above] {\small $\eta$};

\foreach \x in {-1,...,2} {
    \foreach \y in {-1,...,2} {
        \ifnum \x > -1
            \filldraw[darkgray!80] (\x/2,\y/2) circle (1pt);
        \fi
        \ifnum \y > -1
            \filldraw[darkgray!80] (\x/2,\y/2) circle (1pt);
        \fi
    }
}
\filldraw[darkgray!80] (-0.354,-0.354) circle (1pt);
\filldraw[darkgray!80] (-0.354,-0.854) circle (1pt);
\filldraw[darkgray!80] (-0.854,-0.354) circle (1pt);

\end{tikzpicture}
    \caption{
        Schematic of an irregular cell and its 1-ring neighborhood.
        The topology of the initial control mesh is depicted in black
        and the first level of subdivision by the gray dashed lines.
        Based on \cite{Liu_2020ab}.
    }
    \label{fig:cell_irregular}
\end{figure}
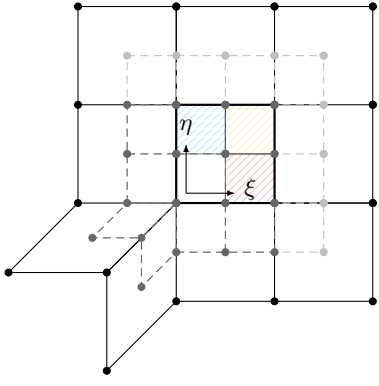

Next, consider the extended subdivision matrices $\vb A$ and $\bar{\vb A}$, as defined in \cite{Stam_1998aa}.
The matrices handle the subdivision process, by mapping initial control points to control points of the refined mesh.
This is illustrated in \cref{fig:cell_irregular}, where the vertices of the initial mesh are shown in black,
and the refined ones in gray.
The dark gray vertices correspond to the matrix $\vb A$ and the light gray ones to $\bar{\vb A}$.

The $2n+8$ local basis functions $\widehat{b}_{\Sigma,i}: [0,1]^2 \to\real$ for an irregular cell $\Sigma$,
collected in a row-vector $\widehat{\vb b}_{\Sigma} = (\widehat{b}_{\Sigma,i})_i$,
are then defined by restriction onto each tile $\Sigma_k^\ell$ as
\begin{equation}
    \widehat{\vb b}_{\Sigma_k^\ell}
    := \qty(\widehat{\vb b}_k \circ t_k^\ell) \overline{\vb A} \vb A^{k-1}.
    \label{eq:catmark_basis_irregular}
\end{equation}
For each $k \in\{1,2,3\}$, the row-vector $\widehat{\vb b}_k: [0,1]^2 \to\real^{2n+17}$ in \eqref{eq:catmark_basis_irregular}
is a collection of the bi-cubic B-splines, sorted to match the ordering of the vertices in the subdivision process
(the light gray ones in \cref{fig:cell_irregular}).
As for the regular case, each irregular basis function $\widehat{b}_{\Sigma,i}$
is associated with one of the $2n+8$ vertices in the neighborhood of the irregular cell.
For a more detailed explanation of the definition and evaluation of the irregular basis functions, see \cite{Stam_1998aa}.

\subsection{Surface parametrization}
Combining the constructions for the regular and irregular basis functions,
a parametrization for the subdivision surface can be given.
Associating a control point $\vb{c}_i \in\real^3$ with each of the $2n+8$ vertices of a cell $\Sigma$,
each local patch of the subdivision surface $S$ can be parametrized by
\begin{equation}
    \Phi: \Sigma\to S
    \quad
    (\xi, \eta) \mapsto \sum_{i=1}^{2n+8} \widehat{b}_{\Sigma,i}(\xi, \eta) \vb{c}_i .
\end{equation}
Note that for the construction of the global parametrization $\Phi:\widehat{S}\to S$,
control points of vertices identified by the neighborhood relation are required to match.

\section{Subdivision-based isogeometric discretization}
\label{sec:iga}

In this chapter, we introduce the discretization of the problem formulation by isogeometric analysis,
i.e. by utilizing the Catmull-Clark basis functions for the discretization of both the domain
and the solution fields.

\subsection{Discrete function spaces}
The Catmull-Clark limit functions, defined on each local cell in \cref{sec:catmark_subd},
are suitable basis functions for an isogeometric Galerkin discretization of the weak forms in \eqref{eq:maxwell_ev_weak_axi}.
On each cell, let
\begin{equation}
    b_{\Sigma,i}: \Phi(\Sigma)\to\real
    \quad
    \vb{p} \mapsto \widehat{b}_{\Sigma,i} (\Phi^{-1}(\vb{p}))
\end{equation}
be the local basis functions on a physical element, assuming that the pullback $\Phi^{-1}$ exists.
For some arbitrary but fixed global ordering of the vertices in $\widehat{S}$ (where vertices identified by the adjacency relation have the same index),
the global subdivision basis functions are defined by association with a vertex each and restriction onto the elements.
Denoting these functions by $b_i: S\to\real$, the subdivision function space is defined as
\begin{equation}
    \mathcal{S}_h := \Span\qty{b_i},
\end{equation}
with $h$ denoting the maximum element size.
For mathematical details on global regularity and linear independence, see \cite{Reif_1995aa, Wawrzinek_2016aa, Peters_2006aa}.

Construction of the subdivision function space in the physical domain
enables the discretization of \eqref{eq:maxwell_ev_weak_axi}.
Consider the discrete representation of the solution $u_h = \sum_i u_i b_i \in\mathcal{S}_h$.
The formulation is then given by
\begin{equation}
    \forall i,j:
    \sum_i u_i \underbrace{(\rot b_i, \rot b_j)_\rho}_{=: k_{ij}}
    = \lambda \sum_i u_i \underbrace{(b_i, b_j)_\rho}_{=: m_{ij}},
    \label{eq:maxwell_ev_iga_axi}
\end{equation}
which yields the generalized eigenvalue problem
\begin{equation}
    \vb K \vb u = \lambda \vb M \vb{u},
\end{equation}
with the stiffness matrix $\vb K = (k_{ij})_{ij}$, mass matrix $\vb M = (m_{ij})_{ij}$
and solution vector $\vb u = (u_i)_i$.

\subsection{Numerical integration}
For the computation of the matrix elements $m_{ij}$ and $k_{ij}$, numerical quadrature is employed
to approximate the value of the integrals with sufficient accuracy.
The standard for FE or B-spline and NURBS based IGA is the use of Gaussian quadrature on every element \cite{Monk_2003aa, Hughes_2005aa}.
Since the basis functions for these methods are smooth in the interior of each element,
Gaussian quadrature has optimal approximation power.
This is however not the case for the subdivision basis functions:
Due to the partitioning of the surface into infinitely many tiles near an extraordinary vertex,
the irregular limit functions are only $C^2$ inside the element.
In order to compute the integrals accurately, a hierarchical quadrature scheme is employed.
The basic idea is, to perform Gaussian quadrature not on the entire element at once,
but on a \emph{finite} number of refined tiles, up until some fixed level $L \geq \ell$.
For more details, see \cite{Juttler_2016aa}.

\section{Numerical results}
\label{sec:results}

The performance of the method proposed in this paper is evaluated in this section,
by numerically solving various model problems.

\subsection{Convergence analysis}
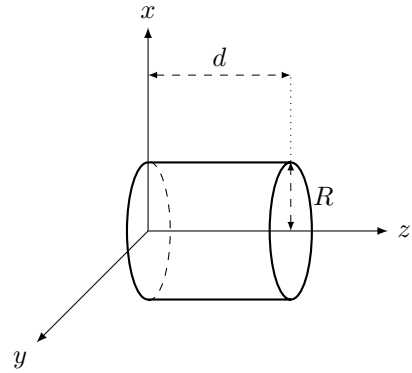
\begin{figure}[ht]
    \centering
\def\height{0.1\textheight}
\def\radius{0.1\textwidth}

\begin{tikzpicture}
    \node[
        cylinder,
        draw=black,
        thick,
        aspect=0.7,
        minimum height=\height,
        minimum width=\radius,
        shape border rotate=0,
        cylinder uses custom fill,
    ] (A) at (0,0) {\phantom{\Huge A}};

    \draw[dashed]
        let \p1 = ($ (A.after bottom) - (A.before bottom) $),
            \n1 = {0.5*veclen(\x1,\y1)},
            \p2 = ($ (A.bottom) - (A.after bottom)!.5!(A.before bottom) $),
            \n2 = {veclen(\x2,\y2)}
     in
        (A.before bottom) arc [start angle=270, end angle=450,
        x radius=\n2, y radius=\n1];

    \path let \p3 = (A.before bottom) in node (origin)  at (\x3,0) {};
    \path let \p4 = (A.before top) in node (top)  at (\x4,0) {};
    \draw[-latex] (origin.center) -- (\height,0) node [pos=1, right] {$z$};
    \node (xaxis) [above = \height of origin] {};
    \draw[-latex] (origin.center) -- (xaxis.center) node [pos=1, above] {$x$};
    \node (yaxis) [below left = 0.71*\height of origin] {};
    \draw[-latex] (origin.center) -- (yaxis.center) node [pos=1, below left] {$y$};

    \node (pt1) [above = \radius of origin] {};
    \node (pt2) [above = \radius of top] {};
    \draw[dotted] (A.before top) -- (pt2.center);
    \draw[latex-latex, very thin, dashed] (pt1.center) -- (pt2.center) node [midway, above] {$d$};
    \draw[latex-latex, very thin, dashed] (top.center) -- (A.before top) node [midway, right, xshift = 0.5em] {$R$};

\end{tikzpicture}
    \caption{
        Shape of the pillbox cavity in Cartesian $(x,y,z)$ coordinates.
        Based on \cite{Ziegler_2023aa}.
    }
    \label{fig:pillbox}
\end{figure}

To numerically verify the discretization scheme, a convergence study is performed.
The first problem to investigate convergence is the cylindrical pillbox cavity,
which has closed-form solutions for the eigenmodes and eigenfrequencies \cite[Chapter 8.7]{Jackson_1998aa}.
A pillbox cavity of radius $R$ and height $d$ is depicted in \cref{fig:pillbox}.
The eigenfrequencies of the transverse magnetic (TM) modes in the cavity are given by
\begin{equation}
    \omega_{mnp} = \frac{1}{\sqrt{\mu\varepsilon}} \sqrt{ \frac{x_{mn}^2}{R^2} + \frac{p^2 \pi^2}{d^2} }
\end{equation}
with $x_{mn}$ being the $n$-th root of the Bessel function $J_m$.

In the following, convergence for the computation of the TM$_{010}$ mode is analyzed,
for the parameters $R = \SI{35}{\milli\meter}$ and $d = \SI{100}{\milli\meter}$.
The eigenfrequency of this mode equals $f \approx\SI{3.278}{\giga\hertz}$
and the fields are given by
\begin{align}
    H_\varphi(\rho,z) &= -i \sqrt{\frac{\varepsilon}{\mu}} E_0 J_1\qty(\frac{x_{01} \rho}{R}) \\
    E_z(\rho,z) &= E_0 J_0\qty(\frac{x_{01} \rho}{R})
\end{align}
with all other components equaling zero. The (non-unique) value $E_0$ denotes the field amplitude.
For the geometry discretization of the axisymmetric slice $[0,R]\times[0,d]$, a purposefully unstructured mesh is utilized,
to highlight the approximation properties under the existence of extraordinary vertices.
Gaussian quadrature of degree $p=4$ is used for the numerical integration on regular elements.
On irregular elements $L=10$ levels of subdivision are used for the hierarchical quadrature scheme.

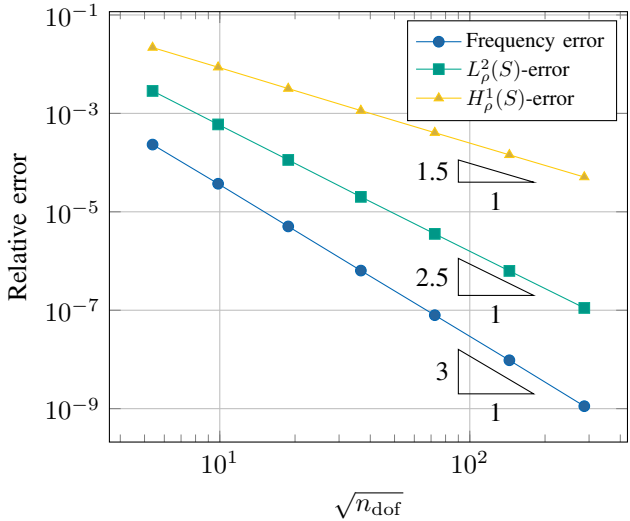
\begin{figure}[ht]
    \centering
    \begin{tikzpicture}[scale=1.0]

\begin{loglogaxis}[
    xlabel=$\sqrt{n_\mathrm{dof}}$,
    ylabel=Relative error,
    grid=major,
    domain=1:100,
    cycle multiindex* list={
        TUDa-b\nextlist
        mark list\nextlist
    },
    legend style={
        inner sep=3pt,
        nodes={scale=0.8, transform shape},
        cells={anchor=west},
    }
] 
    \addplot table[x expr=sqrt(\thisrow{n_dofs}), y=err_f, col sep=comma] 
        {data/pillbox_errs_catmark_irr.csv};

    \addplot table[x expr=sqrt(\thisrow{n_dofs}), y=err_l2, col sep=comma] 
        {data/pillbox_errs_catmark_irr.csv};
 
    \addplot table[x expr=sqrt(\thisrow{n_dofs}), y=err_h1, col sep=comma] 
        {data/pillbox_errs_catmark_irr.csv};

    \draw (90, 11.31e-5) |- (180, 4e-5)
        node[pos=0.25, left] {1.5} node[pos=0.75, below] {1} -- cycle;

    \draw (90, 11.31e-7) |- (180, 2e-7)
        node[pos=0.25, left] {2.5} node[pos=0.75, below] {1} -- cycle;

    \draw (90, 16e-9) |- (180, 2e-9)
        node[pos=0.25, left] {3} node[pos=0.75, below] {1} -- cycle;

    \legend{
        Frequency error,
        $\LtwoAxi[S]$-error,
        $\HgradAxi[S]$-error,
    }
\end{loglogaxis}

\end{tikzpicture}
    \caption{
        Convergence plot for global $h$-refinement of the pillbox cavity test case.
        The relative errors in the eigenfrequency, $\LtwoAxi[S]$-norm and $\HgradAxi[S]$-norm are plotted against the square root $\sqrt{n_\mathrm{dof}}$ of the number of DOFs.
        Both axes are scaled logarithmically.
    }
    \label{fig:conv_plot_pillbox}
\end{figure}

The results of the convergence study using Catmull-Clark elements are depicted in \cref{fig:conv_plot_pillbox}.
The error in the computed eigenfrequency, and the $\LtwoAxi[S]$ and $\HgradAxi[S]$ errors of the eigenmode
are plotted against the square root $\sqrt{n_\mathrm{dof}}$ of the number of degrees of freedom (DOFs).
As indicated by the slope triangles, the observed rates
are approximately \num{1.5} in $\HgradAxi[S]$ and \num{2.5} in $\LtwoAxi[S]$.
These rates agree with observations in the literature for a Catmull-Clark discretization of the $H^1$-Laplace problem,
see e.g. \cite{Liu_2020ab,Dietz_2022aa}.
The study experimentally confirms that the axisymmetric problem exhibits the same convergence rates as the standard 2D problem
and that Catmull-Clark basis functions are hence suitable for axisymmetric discretizations.
The observed rate of \num{3.0} for the eigenfrequency error is also consistent with the literature,
as we would expect a doubling of the $\HgradAxi[S]$-rate \cite[Section 5]{Arnold_2010aa}.

\begin{figure}[ht]
    \centering
\newcommand*{\scale} {1/140mm * \linewidth}
\newcommand*{\req}{\scale * 103.3mm}
\newcommand*{\rir}{\scale * 35mm}
\newcommand*{\rarc}{\scale * 42.0mm}
\newcommand*{\len}{\scale * 57.7mm}
\newcommand*{\mya}{\scale * 12.0mm}
\newcommand*{\myb}{\scale * 19.0mm}

\tikzset{
    partial ellipse/.style args={#1:#2:#3}{
        insert path={+ (#1:#3) arc (#1:#2:#3)}
    }
}

\begin{tikzpicture}[scale=0.8]

\begin{axis}[
    axis lines=left,
    axis equal,
    axis line style={-latex},
    xshift=-1.5cm,
    height=5cm,
    width=10cm,
    ticks=none,
    xlabel={$z$},
    ylabel={$\rho$},
    xlabel style={
        at={(1,0)},
        anchor=west,
    },
    ylabel style={
        at={(0,1)},
        anchor=south,
        rotate=-90,
    },
]

\end{axis}

\draw[very thin] (\len,\req-\rarc) circle (\rarc);
\draw[very thin] (0,\rir + \myb) ellipse ({\mya} and {\myb});

\draw[latex-latex, very thin, dashed] (\len,\req-\rarc) -- ++(130: \rarc) node[pos = 0.4, left] {$R_\mathrm{arc}$};
\draw[latex-latex, very thin, dashed] (-\len*0.05,0) -- (-\len*0.05, \rir) node[midway, left] {$R_\mathrm{iris}$};
\draw[latex-latex, very thin, dashed] (\len*1.05,0) -- (\len*1.05, \req) node[midway, right] {$R_\mathrm{eq}$};
\draw[latex-latex, very thin, dashed] (0,\rir) -- (0, \rir+\myb) node[midway, left] {b};
\draw[latex-latex, very thin, dashed] (0,\rir+\myb) -- (\mya, \rir+\myb) node[midway, above] {a};
\draw[latex-latex, very thin, dashed] (0,-0.05*\rir) -- (\len, -0.05*\rir) node[pos = 0.8, below] {$l$};

\draw[very thin, dotted] (0,\rir+\myb) -- (0, \rir+\myb*2);
\draw[very thin, dotted] (0,\rir+\myb) -- (-\mya, \rir+\myb);

\draw[very thick, TUDa-3c] (0, 0) -- (\len,0);
\draw[very thick, TUDa-3c] (\len,0) -- (\len, \req);
\draw[very thick, TUDa-3c] (\len,\req) arc (90:165:\rarc)  node (node_circ) [pos = 1] {};
\draw[very thick, TUDa-3c] (0,0) -- (0,\rir);
\draw[very thick, TUDa-3c] (0,\rir + \myb) [partial ellipse=-90:-20:{\mya} and {\myb}] node (node_ellip) [pos = 1] {};
\draw[very thick, TUDa-3c] (node_circ.center) -- (node_ellip.center);

\end{tikzpicture}
    \caption{
        Shape of the single-cell TESLA cavity in axisymmetric $(\rho,z)$ coordinates.
        Based on \cite{Ziegler_2023aa}.
    }
    \label{fig:tesla}
\end{figure}
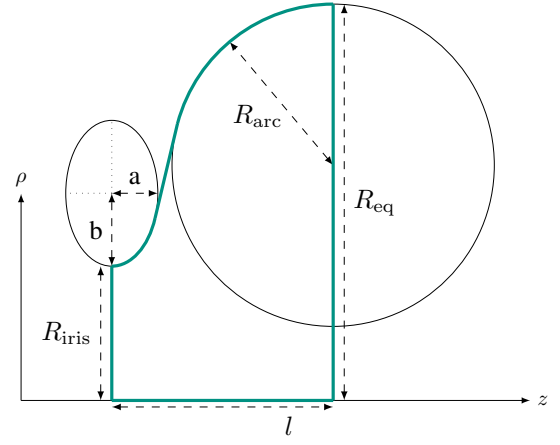

As a more complex test case, we consider the eigenmodes of a single-cell TESLA cavity.
For the experiment, the geometry parameters from \cite{Aune_2000aa} for a midcup are used.
The shape of the cavity is depicted in \cref{fig:tesla}.

In contrast to the pillbox test case, the geometry of the TESLA cavity
is not exactly representable by a Catmull-Clark subdivision surface,
as the 2D cross section contains circular and ellipsoidal arcs.
Therefore, after each refinement step, the boundary of the subdivision surface is fitted to the exact geometry.
For simplicity the fitting is performed by cubic spline quasi-interpolation
\begin{equation}
	\vb{c}_i = \frac{1}{6}\left(-\vb{p}_{i-1} + 8\vb{p}_i - \vb{p}_{i+1} \right) ,
\end{equation}
where $\vb{c}_i$ are the control points of the Catmull-Clark boundary curve
and $\vb{p}_i$ samples of the exact boundary (see \cite{Sablonniere_2005aa}).
More sophisticated methods, like $L^2$-projection or least-squares, are also conceivable,
but are not investigated in this paper.

The convergence of our Catmull-Clark discretization is analyzed for the five lowest TM$_{0np}$ modes.
The utilized quadrature scheme is the same as for the pillbox test case
and the mesh also contains extraordinary vertices.
As there are no closed-form solutions for the eigenfrequencies,
the values used in the error computation are obtained numerically using the open-source IGA framework GeoPDEs \cite{Vazquez_2016aa}
with very high accuracy (using cubic B-splines with $\sim\num{100000}$ DOFs).

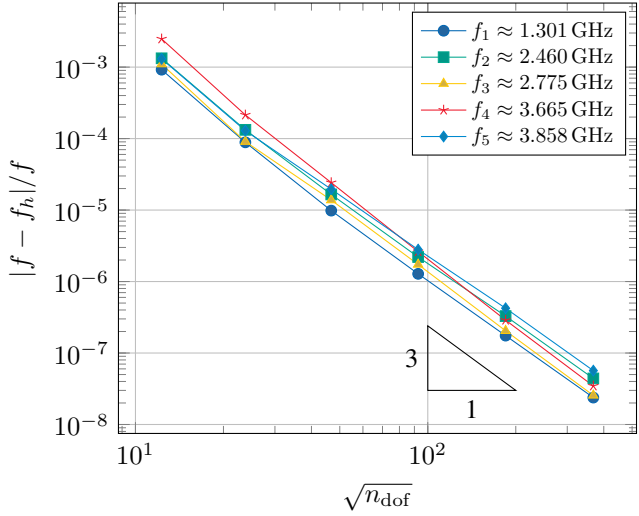
\begin{figure}[ht]
    \centering
    \begin{tikzpicture}[scale=1.0]

\begin{loglogaxis}[
    xlabel=$\sqrt{n_\mathrm{dof}}$,
    ylabel=$\abs{f - f_h} / f$,
    grid=major,
    domain=1:100,
    cycle multiindex* list={
        TUDa-b\nextlist
        mark list\nextlist
    },
    legend style={
        nodes={scale=0.8, transform shape},
        cells={anchor=west},
    }
] 
    \addplot table[x expr=sqrt(\thisrow{n_dofs}), y=err_f1, col sep=comma] 
        {data/tesla_cell_errs_catmark.csv};
    
    \addplot table[x expr=sqrt(\thisrow{n_dofs}), y=err_f2, col sep=comma] 
        {data/tesla_cell_errs_catmark.csv};
    
    \addplot table[x expr=sqrt(\thisrow{n_dofs}), y=err_f3, col sep=comma] 
        {data/tesla_cell_errs_catmark.csv};
    
    \addplot table[x expr=sqrt(\thisrow{n_dofs}), y=err_f4, col sep=comma] 
        {data/tesla_cell_errs_catmark.csv};
    
    \addplot table[x expr=sqrt(\thisrow{n_dofs}), y=err_f5, col sep=comma] 
        {data/tesla_cell_errs_catmark.csv};
    
    \draw (100, 24e-8) |- (200, 3e-8)
        node[pos=0.25, left] {3} node[pos=0.75, below] {1} -- cycle;

    \legend{
        $f_1 \approx \SI{1.301}{\giga\hertz}$,
        $f_2 \approx \SI{2.460}{\giga\hertz}$,
        $f_3 \approx \SI{2.775}{\giga\hertz}$,
        $f_4 \approx \SI{3.665}{\giga\hertz}$,
        $f_5 \approx \SI{3.858}{\giga\hertz}$,
    }
\end{loglogaxis}

\end{tikzpicture}
    \caption{
        Convergence plot for global $h$-refinement of the TESLA single-cell test case.
        The relative eigenfrequency error for the first five modes is plotted against the square root $\sqrt{n_\mathrm{dof}}$ of the number of DOFs.
        Both axes are scaled logarithmically.
    }
    \label{fig:conv_plot_tesla_cell}
\end{figure}

The results are shown in \cref{fig:conv_plot_tesla_cell}.
For all analyzed eigenmodes, the trend for the frequency error decay is roughly the same,
differing only in a constant.
Similar to the pillbox test case, the observed rates are all approximately \num{3},
validating the applicability of the method for complex geometries.

\subsection{Longitudinal field evaluation}
\begin{figure}[ht]
    \centering
    \begin{tikzpicture}[
    scale=1.0,
    spy using outlines={
        circle,
        magnification=5,
        size=2cm,
        connect spies,
    },
]

\begin{axis}[
    xlabel=$z$,
    ylabel=$E_\rho$,
    use units,
    x unit=\meter,
    y unit=\volt\squared\per\meter\squared,
    xmin=0.02,
    xmax=1,
    grid=major,
    cycle list name=TUDa-b,
    legend style={
        inner sep=3pt,
        nodes={scale=0.8, transform shape},
        cells={anchor=west},
    },
    legend image post style={
        line width=1pt,
    }
]
    \addplot+[line width=0.15] table[x=z, y=e_r, col sep=comma]
        {data/eval_tesla_mid_equator_lin.csv};

    \addplot+[line width=0.15] table[x=z, y=e_r, col sep=comma]
        {data/eval_tesla_mid_equator_catmark.csv};

    \coordinate (target) at (axis cs:0.14,-155);
    \coordinate (zoom) at (axis cs:0.7,-100);
    \spy on (target) in node[fill=lightgray!50] at (zoom);

    \legend{
        Lowest order FE,
        Catmull-Clark
    }
\end{axis}

\end{tikzpicture}
    \caption{
        Graph of the $\rho$-component of the numerical electric field in the 9-cell TESLA cavity.
        Evaluation is performed along the horizontal line $\rho = \SI{51,65}{\milli\meter}$.
    }
    \label{fig:eval_fields_tesla_mid_equator}
\end{figure}
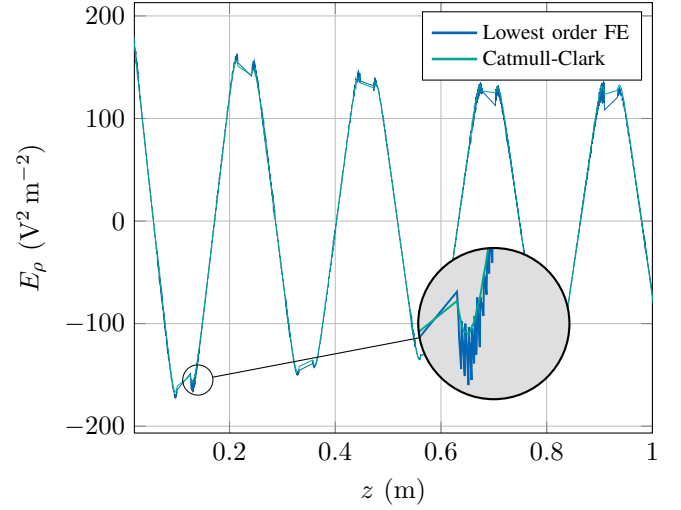

As an illustration of the increased smoothness of the eigenmodes,
we showcase the graph of the electric field inside a 9-cell TESLA cavity with respect to the longitudinal $z$-direction.
The geometry parameters for the cavity cells are again taken from \cite{Aune_2000aa}.
The radial component $E_\rho$ of the field is evaluated
along the horizontal lines $\rho = R_\text{iris}/2$ and $\rho = R_\text{eq}/2$
and plotted in \cref{fig:eval_fields_tesla_mid_iris} and \cref{fig:eval_fields_tesla_mid_equator}, respectively.
The plots showcase a comparison between lowest order FE and Catmull-Clark elements.
Both discretizations use the same mesh with extraordinary vertices.

The results visually confirm that the Catmull-Clark discretization method produces accurate results for the electric field,
as the solution qualitatively agrees with the standard finite element one.
However, it can also be seen, that the solutions noticeably differ towards the peaks of the field:
For lowest order $C^0$ elements, the electric field becomes increasingly noisy,
while the Catmull-Clark solution does not show overshoots.
These unphysical jumps at element interfaces can be attributed to the discretization method,
since $C^0$ basis functions cannot guarantee a continuous solution for the derivative (here, the electric field).
Because Catmull-Clark basis functions are $C^2$ across elements, they guarantee continuity even of the derivatives.
Using higher order $C^0$ basis functions (e.g. Lagrange elements) is also expected to significantly suppress the observed numerical noise,
resulting in a better field quality.
However, this approach cannot exactly suppress the jumps in the derivative
and requires more degrees of freedom on the same mesh.

\begin{figure}[ht]
    \centering
    \begin{tikzpicture}[
    scale=1.0,
    spy using outlines={
        circle,
        magnification=5,
        size=2cm,
        connect spies,
    },
]

\begin{axis}[
    xlabel=$z$,
    ylabel=$E_\rho$,
    use units,
    x unit=\meter,
    y unit=\volt\squared\per\meter\squared,
    xmin=-0.1,
    xmax=1.1,
    grid=major,
    cycle list name=TUDa-b,
    legend style={
        inner sep=3pt,
        nodes={scale=0.8, transform shape},
        cells={anchor=west},
    },
    legend image post style={
        line width=1pt,
    }
]
    \addplot+[line width=0.15, each nth point=2] table[x=z, y=e_r, col sep=comma]
        {data/eval_tesla_mid_iris_lin.csv};

    \addplot+[line width=0.15, each nth point=2] table[x=z, y=e_r, col sep=comma]
        {data/eval_tesla_mid_iris_catmark.csv};

    \coordinate (target) at (axis cs:0.115,280);
    \coordinate (zoom) at (axis cs:0.7,-100);
    \spy on (target) in node[fill=lightgray!50] at (zoom);

    \legend{
        Lowest order FE,
        Catmull-Clark
    }
\end{axis}

\end{tikzpicture}
    \caption{
        Graph of the $\rho$-component of the numerical electric field in the 9-cell TESLA cavity.
        Evaluation is performed along the horizontal line $\rho = \SI{17.5}{\milli\meter}$.
    }
    \label{fig:eval_fields_tesla_mid_iris}
\end{figure}
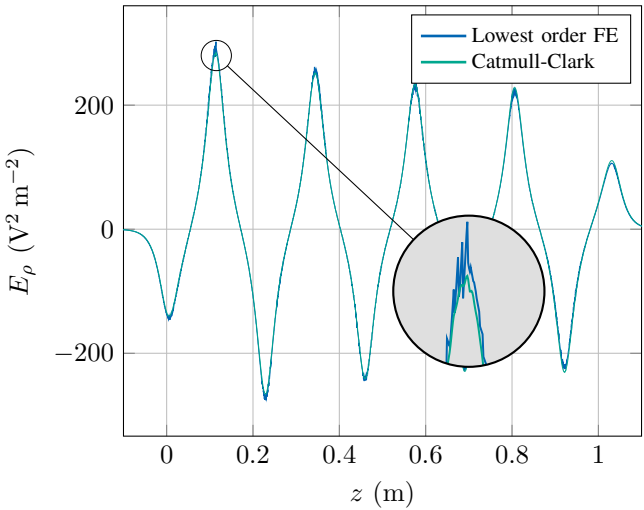

\section{Conclusion}
\label{sec:conclusion}

In this work, we have presented a subdivision-based isogeometric discretization of the axisymmetric Maxwell eigenvalue problem.
The discretization is based on the Catmull-Clark subdivision scheme, utilizing the subdivision basis functions for both the representation of the geometry
and for the discretization of the axisymmetric unknowns.
For TM modes, the resulting magnetic field is $C^2$-continuous except at extraordinary vertices, and the electric field is $C^1$-continuous.

Performance of the method has been demonstrated by various numerical experiments.
A convergence analysis for the computation of TM modes in a pillbox cavity showed
that the rates in $L^2$ and $H^1$ agree with observed rates for the two-dimensional Poisson problem in the literature
(\num{2.5} and \num{1.5} respectively).
The eigenvalues of the pillbox and single-cell TESLA cavity also converged with the expected (doubled) rate.

The quality in terms of smoothness of the numerical fields of a 9-cell TESLA cavity
has been investigated and compared to classical lowest order finite elements.
The results show that the subdivision-based discretization yields significantly smoother fields,
overcoming the interface discontinuities of the finite element solution,
while requiring the same amount of degrees of freedom.

\bibliographystyle{IEEEtranN}
\bibliography{bibtex}

\end{document}